\documentclass[epsfig,11pt]{article}
\usepackage{epsfig,citesort,amssymb,amsmath}

      \input{amssym.def}
      \input{amssym}
      \newcommand {\mm}[1] {\ifmmode{#1}\else{\mbox{\(#1\)}}\fi}

\newcommand{\utwi}[1]{\mbox{\boldmath $ #1$}}

\newcommand{\bx}{{\utwi{x}}}

\newcommand{\bP}{{\utwi{P}}}
\newcommand{\bQ}{{\utwi{Q}}}

\newcommand{\bT}{{\utwi{T}}}

\begin{document}
      \title{\bf
Are residues in a protein folding nucleus
evolutionarily conserved?
}

      \author{\bf Yan Yuan Tseng and Jie Liang\thanks{Corresponding author. Phone:
      (312)355--1789, fax: (312)996--5921, email: {\tt
      jliang@uic.edu}} \\ Department of Bioengineering, SEO, MC-063 \\
      University of Illinois at Chicago\\ 851 S.\ Morgan Street, Room
      218 \\ Chicago, IL 60607--7052, U.S.A.\\
{(Accepted by {\it J. Mol. Biol.})}
}  \date{\today}

\maketitle
\vspace*{0.5in} 

\abstract{ Protein is the working molecule of cell, and evolution is
the hallmark of life.  It is important to understand how protein
folding and evolution influences each other.  Several studies
correlating experimental measurement of residue participation in
folding nucleus and sequence conservation have reached different
conclusions.  These studies are based on assessment of sequence
conservation at folding nucleus sites using entropy or relative
entropy measurement derived from multiple sequence alignment.  Here we
report analysis of conservation of folding nucleus using an
evolutionary model alternative to entropy based approaches.  We employ
a continuous time Markov model of codon substitution to distinguish
mutation fixed by evolution and mutation fixed by chance.  This model
takes into account bias in codon frequency, bias favoring transition
over transversion, as well as explicit phylogenetic information.  We
measure selection pressure using the ratio $\omega$ of synonymous {\it
vs.}\ non-synonymous substitution at individual residue site.  The
$\omega$-values are estimated using the {\sc Paml} method, a
maximum-likelihood estimator.  Our results show that there is little
correlation between the extent of kinetic participation in protein
folding nucleus as measured by experimental $\phi$-value and selection
pressure as measured by $\omega$-value.  In addition, two
randomization tests failed to show that folding nucleus residues are
significantly more conserved than the whole protein, or the median
$\omega$ value of all residues in the protein.  These results suggest
that at the level of codon substitution, there is no indication that
folding nucleus residues are significantly more conserved than other
residues.  We further reconstruct candidate ancestral residues of the
folding nucleus and suggest possible test tube mutation studies for
testing folding behavior of ancient folding nucleus.  }
\vspace*{0.7in}


\newpage

Are amino acid residues important for rapid folding preferentially
conserved during evolution?  Does natural selection optimize proteins
for folding kinetics?  If protein folding involves initially the
formation of a small region of native-like folding nucleus, are
identities of these residues well conserved during evolution
\cite{Shrivastava95_PNAS,Shakh96_Nature,Mirny98_PNAS,Ptitsyn98_JMB,MichnickShakhnovich98_FD}?
These fundamental questions of molecular biology have received much
attention
\cite{Shrivastava95_PNAS,Shakh96_Nature,Mirny98_PNAS,Ptitsyn98_JMB,MichnickShakhnovich98_FD,PtitsynTing99_JMB,MirnyShakh01_JMB,Plaxco00_JMB,Larson02_JMB,Grantcharova98_NSB,Fulton00_JMB,Demirel98_PS}.
Of direct relevance are experimental $\phi$-value studies, which
provide information about the role individual amino acid residue play
in the formation of folding nucleus
\cite{Fersht97_COSB,Matouschek89_Nature,Matouschek91_ME}.  By
measuring the change $\Delta \Delta G$ in protein stability and the
change $\Delta \Delta G^{\ddagger}$ in folding barrier due to mutation
of an amino acid residue, $\phi$-value (defined as $ \phi \equiv
\Delta \Delta G^{\ddagger}/\Delta \Delta G $) for the mutated residue
can be calculated.  $\phi$-value has been used to measure the extent
to which side chain of a mutated residue participates in native-like
interactions. A $\phi$-value of $0.0$ indicates that the site of
mutation is as unfolded as in the denatured state.  A $\phi$-value of
$1.0$ indicates that the site of mutation is as folded as in the
native state, {\it i.e.}, this residue is involved in native-like
transition state structure, and is part of the folding
nucleus. $\phi$-value between 0 and 1 is interpreted as possessing
different degrees of structure in transition state
\cite{Fersht97_COSB}.  The folding nucleus can be identified as formed
by the set of residues with $\phi$ values above a threshold ({\it
e.g.}, $\phi \ge 0.5$) \cite{Fersht97_COSB}.  Several computational
methods have been developed for predicting protein folding mechanism and
$\phi$-values of residues.  These include the sequential binary
collision model \cite{AlmBaker99_PNAS}, multisegment model
\cite{GalzitskayaFinkelstein99_PNAS}, and single-to-triple sequence
approximation model \cite{MunozEaton99_PNAS}.  Model conformations of
transition-state ensemble have also been generated explicitly by Monte
Carlo sampling using  G\={o}-type potential derived from experimental
$\phi$-values constraints \cite{Vendruscolo01_Nature}.  A lucid
statistical mechanistic picture for understanding $\phi$-value
experiments can be found in \cite{Ozkan1_NSB,Ozkan02_PS}.

The evolutionary conservation of folding nucleus residues are the
subject of several recent studies.  These studies, however, have come
to different conclusions.  Plaxco {\it et al\/} and Larson {\it et
al\/} showed that there may be little correlation between sequence
conservation and participation in the folding transition state
\cite{Plaxco00_JMB,Larson02_JMB}.  Mirny, Shahknovich and others
demonstrated that for rapid folding, sequence identity of folding
nucleus are more conserved within protein families and across protein
superfamilies \cite{Shakh96_Nature,MirnyShakh01_JMB}.  
It is unclear
whether the disagreement between these studies is due to the
difference in entropy calculations as attributed in
\cite{MirnyShakh01_JMB}, or differences in choice and processing of
the data set, in sequence alignments, in definition of folding
nucleus, as well as intrinsic sample bias in $\phi$-value analysis, as
discussed in details in \cite{Larson02_JMB}.

In this study, we examine the conservation of folding nucleus residues
using an approach that differs from previous studies in several
aspects.  First, instead of studying amino acid residue sequences, we
examine the evolution of corresponding coding DNA sequences at codon
level.  Second, we use an explicit codon evolutionary model based on
continuous time Markov process, which has yielded deep insights about
the mechanisms of molecular evolution
\cite{Yang01_HSG,Swofford96_book,LioGoldman98_GR}.  Instead of using
entropy or relative entropy as quantitative measure of sequence
conservation, we assess the ratio of mutation rates of synonymous {\it
vs.}\ non-synonymous changes to detect natural selection at each amino
acid residue position.  Third, a phylogenetic tree is built to encode
the closeness between proteins.  Following previous studies
\cite{GoldmanYang94_MBE,NielsenYang98_Genetics}, we use maximum
likelihood method developed in \cite{PAML} by Yang to estimate values
of parameters of the evolutionary model and draw inference about the
conservation of folding nucleus residues.

We find experimental $\phi$-values are not correlated with
evolutionary conservation for seven proteins studied here.  In
addition, results using two statistical tests indicate that except
possibly one protein, none of these proteins have folding nucleus more
conserved than the rest of the proteins, or than the residue with
median selection pressure.  We have also reconstructed candidate
ancestral folding nucleus residues, and have suggested exploratory test-tube
mutation studies on the evolution of protein folding dynamics.

{\bf Synonymous and nonsynonymous codon substitution.}  Protein
sequences diverge from a common ancestor because mutations occur. Some
fraction of these mutations are fixed into the evolving population by
selection and some are fixed by chance, resulting in the substitution
of one nucleotide for another nucleotide at various locations.
Because evolution occurs at DNA level rather than at amino acid level,
models of protein evolution based on codon usage are appealing and
have been widely used
\cite{Schoniger90_JTB,GoldmanYang94_MBE,MuseGaut94_MBE,Yang98_MBE}.
In this study, we therefore consider substitutions at the codon
level. A codon substitution can have two different outcomes for the
nucleotide sequence of protein coding region: {\it synonymous
substitution\/} does not change the encoded sequence of amino acid
residues, whereas {\it nonsynonymous substitution\/} leads to changes
in the amino acid residues.  Random mutation and selection pressure
will have different effects on the rate of these two types of
substitutions \cite{Kimura83,Gillespie94,NeiGojobory86_MBE}, and this
difference can be exploited for detecting selective pressure at
protein level
\cite{HughesNei88_Nature,Li93_JME,MessierStewart97_Nature,Yang98likelihood_MBE,GoldmanYang94_MBE,YangNielsen98_JME}.
Our key problem is to find out the ratio of the synonymous
substitution rate $d_s$ and the nonsynonymous substitution rate $d_n$.
That is, we wish to estimate the ratio of the numbers of synonymous
and non-synonymous substitutions at a specific site or a specific
amino acid residue position.  If natural selection offers no
advantage, non-synonymous mutations will have the same rate as
synonymous mutations ($d_n = d_s$), and the ratio $\omega = d_n/d_s$
will be 1.  If non-synonymous mutations are harmful, deleterious or
lethal, purifying selection is at play and the rate for non-synonymous
mutation will be reduced: we have $d_n < d_s$ and $\omega <1$.  On the
other hand, if Darwinian positive selection favors non-synonymous
mutation, we have $d_n > d_s$ and $\omega >1$.  Here $\omega$ is used
as a measure of selection pressure.  Substitution fixed by evolution
and substitution fixed by chance are distinguished by examining the
ratio $\omega$ at various locations of amino acid residues.  This
technique has been frequently applied in studies of molecular
evolution, {\it e.g.}, in detecting adaptive evolution
\cite{Yang01_HSG,YangNielsen98_JME,Swanson01_PNAS}.

{\bf Continuous time Markov process for codon substitution.}  Markov
model has been widely used in sequence analysis \cite{DEKM98} and in
evolutionary models \cite{Swofford96_book}.  In the current model, the
outcome of codon substitution is determined only by the identity of
codon in the ancestral sequence separated by divergence time $t$, and
a codon transition probability matrix ${\bf P}(t)$.  A phylogenetic
tree is a key ingredient of this model.  The topology and branch
lengths of the tree reflects the evolutionary relationship among
different proteins, which can model their closeness
\cite{Swofford96_book}.  We follow the approach of
\cite{NielsenYang98_Genetics,Yang00_Genetics,Yang01_HSG}, and describe
below briefly the model.

For a given phylogenetic tree, the parameters of the evolutionary
model are a $61 \times 61$ rate matrix $\bQ$ for 61 non-stop codons
and the sequence divergence time $t$s (or the branch lengths) of the
phylogenetic tree. The divergence time represents expected number of
changes between sequences which are nodes in a phylogenetic tree.  The
entries $q_{ij}$ of matrix $\bQ$ are infinitesimal substitution rates
of nucleotides for the set ${\cal C}$ of 61 non-stop codons, and they
are parametrized as:
\[
q_{ij} = \left \{
\begin{array}{ll}
0,            & \mbox{if $i$ and $j$ differ at two or three codon positions,}\\
\mu \pi_j,    & \mbox{if $i$ and $j$ differ by a synonymous transversion,}  \\
\mu \kappa \pi_j,    & \mbox{if $i$ and $j$ differ by a synonymous transition,}  \\
\mu \omega \pi_j,    & \mbox{if $i$ and $j$ differ by a non-synonymous transversion,}  \\
\mu \omega \kappa \pi_j,    & \mbox{if $i$ and $j$ differ by a non-synonymous transition,}
\end{array}
\right.
\]
where $\mu$ is the basis rate, $\kappa$ is the transition/transversion
rate ratio, $\omega$ the ratio of nonsynonymous and synonymous rates,
and $\pi_j$ is the codon frequency, which can be estimated as observed
codon frequency in the sequences.  In this model, the $61 \times 61$
rate matrix ${\bf Q}$ is fully determined by two parameters $\kappa$ and $\omega$,
since $\pi_j$ can be estimated and $\mu$ is a constant
\cite{GoldmanYang94_MBE,NielsenYang98_Genetics}.

For continuous time Markov process, the transition probability matrix
of size $61\times 61$ after time $t$ is \cite{LioGoldman98_GR}:
\[
\bP(t) = \{ p_{ij}(t) \} = \exp (\bQ \cdot t) 
\]
The entry $p_{ij}(t)$ represents the probability that codon $i$ will
mutate into codon $j$ after time $t$.  It is calculated through
diagonalization of the $\bQ$ matrix.

{\bf $\omega$ ratio from likelihood of  phylogeny.}  For node $i$
and node $j$ in a phylogenetic tree separated by divergence time
$t_{ij}$, the time reversible probability of observing nucleotide
$x_i$ in a position $h$ at node $i$ and nucleotide $x_j$ of the same
position at node $j$ is:
\begin{equation}
\pi_{x_i} p_{x_i x_j}(t_{ij}) = \pi_{x_j} p_{x_j x_i}(t_{ij}).
\label{reversible}
\end{equation}
For a set ${\cal S}$ of $s$ multiple-aligned sequences with $n$ amino
acid residues, we assume that a reasonably accurate phylogenetic tree
$\bT = ({\cal V}, {\cal E})$ is given. Here $\cal V$ is the set of
nodes (or vertices), namely, the union of the set of observed $s$
sequences $\cal L$ (leaf nodes), and the set of $s-2$ ancestral
sequences $\cal I$ (internal nodes). $\cal E$ is the set of edges (or
branches) of the tree.  Let the vector $\bx_h = (x_1, \cdots, x_s)^T$
be the observed codons at position $h$ for the $s$ sequences. Without
loss of generality, we assume that the root of the phylogenetic tree
is an internal node $k$.  Given the specified topology of the
phylogenetic tree $\bf T$ and the set of branch lengths (or divergence
times), and if the set of codons ${\cal C_{\cal I}}$ of all internal
nodes {$\cal I$} is specified, the probability of observing the $s$
number of codons $\bx_h$ at position $h$ is:
\[
p(\bx_h|{\cal C}_{\cal I}, \bT) =  \pi_{x_k} \prod_{(i,j) \in {\cal E}} p_{x_i x_j}(t_{ij}).
\]
Summing over the set ${\cal C}$ of all possible codons for the
internal nodes ${\cal I}$, we have
\begin{equation}
p(\bx_h| \bT) =
\pi_k
\sum_{\substack{
{i \in {\cal I}}\\
{x_i \in {\cal C}}
} }
\prod_{(i,j) \in {\cal E}}
p_{x_i x_j}(t_{ij}).
\label{sameRate}
\end{equation}
The probability of
observing all codons in the coding region of the nucleotide sequences
is:
\[
P({\cal S}|\bT ) =  P(\bx_1, \cdots, \bx_s|\bT) =\prod_{h=1}^s p(\bx_h|\bT).
\]

To account for the possibility that the rate of nonsynonymous
substitution can vary among different sites, the model developed in
\cite{Yang00_Genetics} allows $M$ possible different classes of
nonsynonymous substitutions with rates $\omega_1, \cdots,
\omega_M$. Each amino acid site falls into the $M$ class with
probabilities $p_1, \cdots, p_M$ \cite{Yang00_Genetics}.  The
probability of observing $\bx_h$ is then modified from Equation
(\ref{sameRate}), which gives $p(\bx_h|\omega_m, \bT)$, to the
following:
\[
p(\bx_h|\bT) = \sum_{m=1}^M p_m \cdot p(\bx_h|\omega_m, \bT).
\]
Repeat this calculation over all amino acid residue sites, we have
\[
P({\cal S}|\bT) =
\prod_{h=1}^s p(\bx_h|\bT),
\]
and the likelihood function is:
\[
\ell (\bT)
)
= \sum_{h=1}^s \log[p(\bx_h|\bT)].
\]

To estimate the parameters $\kappa_h$, $\omega_h$ for each site $h$
used in the mutation rate matrix $\bQ$, we use a Maximum Likelihood
Estimator
\cite{Felsenstein81_JME,NielsenYang98_Genetics,Yang98likelihood_MBE},
the {\sc Paml} package by Yang \cite{PAML}.  Our goal is to search for
parameters $ \kappa_h$ and $\omega_h$ such that the likelihood
function $\ell(\bT)$ is maximized. Here the number $M$ of different
classes of $\omega$ is $10$, and they take the default values as
assigned by {\sc Paml} \cite{Yang00_Genetics}.

Once the model parameters are estimated, the empirical Bayes approach
can be used to infer the most likely class of $\omega$ value at each
residue site \cite{Yang01_HSG}.  In {\sc Paml}, the posterior
probability 
$ p(\omega_m|\bx_h)$
that site $h$ with observed codons $\bx_h$ is from class
$m$ with  rate ratio $\omega_m$ is calculated as: 
\[
 p(\omega_m|\bx_h) = p_m \cdot
p(\bx_h|\omega_m, \bT)/p(\bx_h|\bT) = p_m \cdot p(\bx_h|\omega_m,
\bT)/\sum_m p_m  \cdot p(\bx_h|\omega_m, \bT). 
\]

\begin{figure}[h!]
\centerline{
      \epsfig{figure=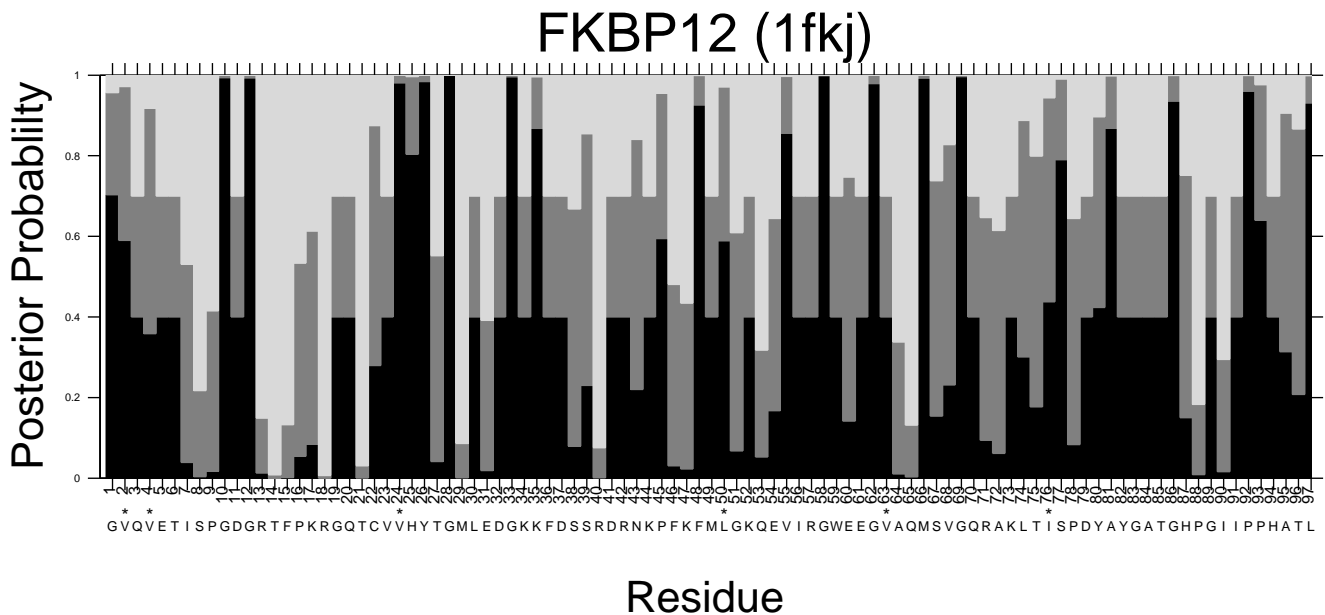,width=7in}
}
\caption{\small \sf
Selection pressure as
measured by $\omega$ ratio of nonsynonymous {\it vs.}\ synonymous
codon substitution rate varies at each amino acid residue site along the
sequence of protein FKBP12.  The ten possible $\omega$ values are
grouped into three classes: $\omega_a< 0.12$ (dark), $0.12\le \omega_b
<0.34$ (gray), and $0.34 < \omega_c$ (light).  The $x$-axis shows the
residue number of the protein, the $y$-axis shows the posterior
probability of $\omega$ belonging to one of the three classes at each
codon position.  Residues with large probability for $\omega_a$ (dark)
are highly conserved residues experiencing strong purifying pressure.
Folding nucleus residues as identified in \cite{MirnyShakh01_JMB} are
marked by the symbol ``*''.
}
\label{Fig:omega-bar}
\end{figure}

{\bf Data collection and computational procedures.}  We follow
\cite{MirnyShakh01_JMB} and study evolution of the set of proteins
taken from Table 1 of \cite{MirnyShakh01_JMB}, where the folding
nucleus residues are defined.  We first query with the sequence of each
of the proteins against HSSP database \cite{HSSP} to obtain homologous
protein sequences with overall sequence identity $>30\%$ to ensure
that they have the same fold.  In some cases, we also searched the
{\sc Ce} server \cite{CE} for structural homologs.  Experimentation
using {\sc Psi-blast} searching the NR-database of protein sequences
give almost identical sets of sequences.  In this study, all redundant
sequences are removed. Since paralogous sequences in a single species
may exist that can be matched to the query DNA sequence, we only take
the sequence with the highest identity to the query protein when
multiple homologous sequences are found in a single species.  With the
exception of protein CI2 where sequences of two paralogs are
included, only proteins with $ \ge 5$ known orthologous DNA sequences
are kept. We therefore exclude AcP protein and CD2.d1 protein because
fewer than 5 DNA sequences were found.  Since
paralogs are excluded, the number of sequences used in this study is smaller than
that used in other studies
\cite{Plaxco00_JMB,MirnyShakh01_JMB,Larson02_JMB}.  The amino acid
residue sequences of the remaining 7 proteins are first aligned using
{\sc ClustalW} with default parameters \cite{ClustalW} and then with
manual intervention.  Alignment of the nucleotide sequences are
generated following the alignment of the protein sequences.  A
phylogenetic tree ${\bf T}$ is constructed using maximum likelihood
method as implemented in the {\sc Paup} method \cite{Paup}.  This tree
${\bf T}$ is then used by the {\sc Paml} package, an implementation of
the maximum likelihood method for estimating $\omega$ values
\cite{PAML}.  In many cases, minor difference in the tree does not
affect final results significantly
\cite{Ford01_MBE,YangSwanson02_MBE}.  For each protein, we repeatedly
estimate $\omega$ twenty times using different initial $\omega$ value
that is assigned to all amino acid sites.  The initial $\omega$ values
range from $0.01$ to $2.00$, at an interval of $0.1$.  About $90\%$ of
the computation converges.  For each protein, all different converged
estimations among the twenty calculations give identical $\omega$
parameters at individual codon positions.

{\bf Natural selection at protein folding nucleus.}  The estimation of
site-specific $\omega$-values can uncover residues important for
biological function, for structural stability, and potentially for
folding kinetics.  In this study we focus on the natural selection of
folding nucleus residues which are identified by $\phi$-value
experiments.  An example for estimated $\omega$ values is shown in
Fig~\ref{Fig:omega-bar}.

\begin{figure}[t!]
      \centerline{
     \epsfig{figure=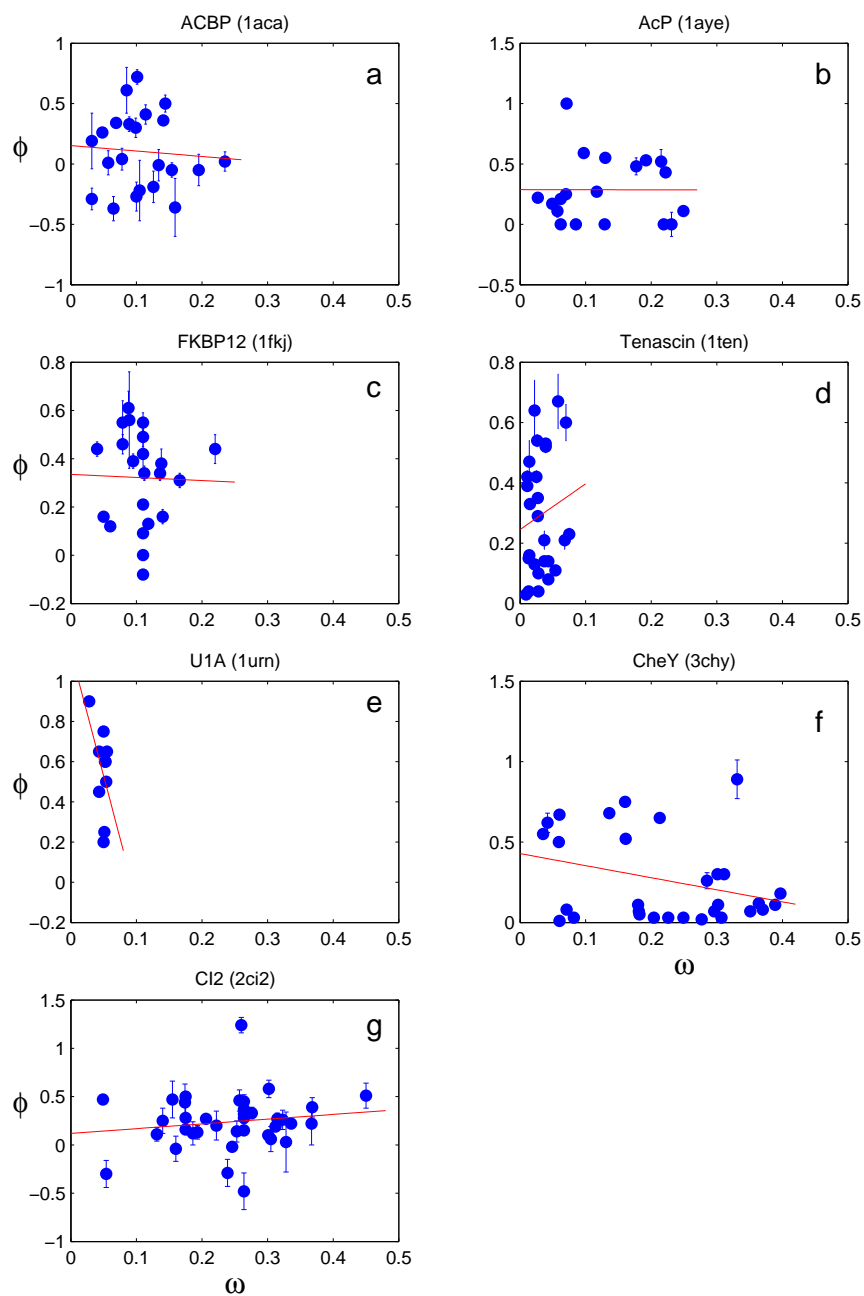,width=4.5in}
}
\caption{\small \sf
Participation in folding
nucleus as measured by experimental $\phi$-value and selective
pressure as measured by $\omega$-value are poorly correlated.
}
\label{Fig:phi-omega}
\end{figure}

We first examine the patterns of $\omega\/$-ratio of nonsynonymous
{\it vs.}\ synonymous substitutions in the seven proteins.  If folding
nucleus residues are more conserved than other residues, selection
pressure then must be correlated with the extent of participation in
folding nucleus \cite{Larson02_JMB}.  Following Larson {\it et al}, we
examine directly the correlation of the $\phi$-values and the
$\omega$-values of characterized residues for each protein.  This
approach helps to circumvent the unavoidable arbitrariness in the
assignment of the set of folding nucleus residues
\cite{Larson02_JMB,Fersht97_COSB}.  Residues with characterized
$\phi$-values for these proteins are obtained from references cited in
\cite{Larson02_JMB}.  Following \cite{Plaxco00_JMB}, we exclude
residues with $\phi<-0.5$ or $\phi>1.5$, and require all $\phi$-values
to have standard deviation $<1.0$, with the exception of protein U1A
({\tt 1urn}), where no data of standard deviations are provided.

\begin{figure}[!t]
\centerline{
      \epsfig{figure=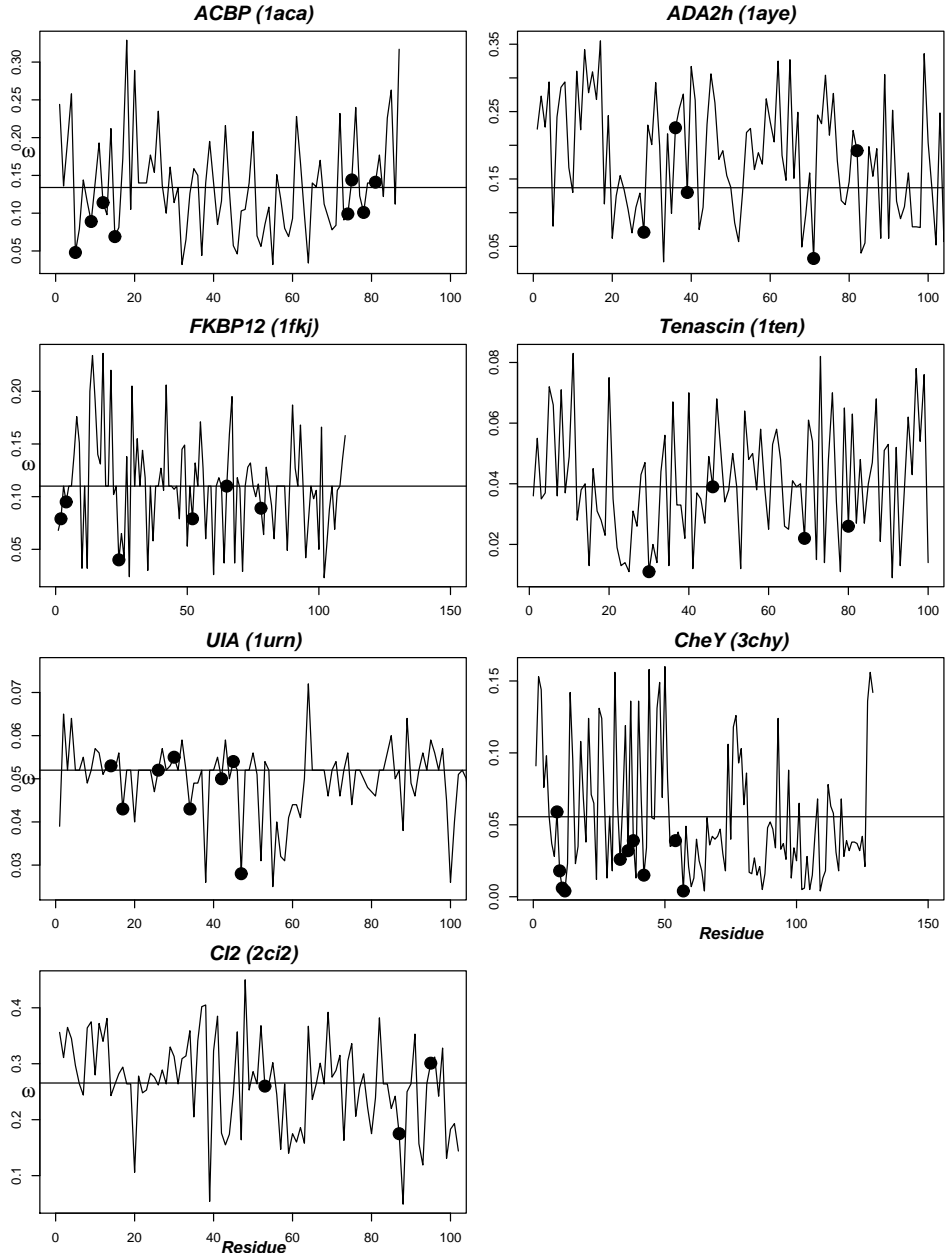,width=5in}
}
\caption{
\small \sf
The weighted mean value
$\bar{\omega} = \sum_{m=1}^{10} p_m \cdot \omega_m$ of estimated
$\omega$ ratio at each residue position of the proteins.  The
$x$-axis shows the residue number of the protein, the $y$-axis shows
the estimated $\bar{\omega}$ at each residue position.  The horizontal
line marks the median $\bar{\omega}$ value of all positions.  Folding
nucleus residues as identified in \cite{MirnyShakh01_JMB} are marked
by ``*''.  Except protein CheY, randomization tests show that folding
nucleus residues are not more conserved than the rest of the protein,
and in all protein cases (including CheY protein) are not more
conserved than the residue at 50\% quantile of all residues ranked by
$\omega$.
}
\label{Fig:omega}
\end{figure}

Among the set of residues with experimentally characterized
$\phi$-values, there is little correlation between $\phi$-value and
$\omega$-value (Fig~\ref{Fig:phi-omega}). The $R^2$ values range
between 0.0 to 0.22, and the two-sided $p$-values of $t$-test for the
null hypothesis that the slope of the linear regression models is 0
range from 9\% to 99\% (Table~\ref{Tab:random-test}). That is,
there is no indication of significant correlation between the extent
of kinetic participation as measured by $\phi$-value and selection
pressure as measured by $\omega$-value.  Our results are similar to
those found in \cite{Plaxco00_JMB,Larson02_JMB}, where relative
entropy instead of $\omega$ was used as the measure of evolutionary
conservation.

The weighted mean values of estimated $\omega$ ratio $\bar{\omega} =
\sum_m p_m \cdot \omega_m $ at each codon position are plotted in
Figure~\ref{Fig:omega}.  It is clear that for each protein, many
folding nucleus residues as defined in \cite{MirnyShakh01_JMB} have
small values of $\omega$, many are often smaller than the median
$\omega$-value of all codon positions.  This indicates that folding
nucleus residues experience purifying selection pressure. However,
there are also many other residues with small $\omega$-value, some of
which have not be characterized by $\phi$-value studies. As discussed
in \cite{Larson02_JMB}, the lower $\omega$-values of folding nucleus
as defined in \cite{MirnyShakh01_JMB} residues could also be a
reflection of the experimental bias in choosing conserved protein core
residues for $\phi$-value experiments.  Can we still conclude that
experimentally identified folding nucleus residues in general are more
conserved than the rest of the protein?

\begin{table*}[t!]
\begin{center}
\begin{tabular}{|lll||c c||c c||ll||ll|}
\hline
Protein & PDB     &$N_{\mbox{prot}}$ &$N_{\mbox{seq}}$ & $N_{\phi}$  & $^a R^{2}$  & $^a p$ & $^b p_{\mbox{all}}$  & $^b p_{\mbox{50\%}}$ &  $^c Z_{\alpha,\mbox{fn}}$ & $^c Z_{\alpha, \mbox{all}}$\\
\hline \hline
CI2     &  2ci2I & 83    & 5 & 37 & $2.1\times 10^{-2}$ &0.39 & $3.3\times10^{-1}$  & $8.6\times 10^{-1}$   &3.29& 2.81 \\
Tenascin &  1ten & 2201  & 5 & 27 & $2.1\times 10^{-2}$ &0.47 & $4.1\times 10^{-2}$ & $2.5\times 10^{-1}$   & 3.29& 3.44 \\
CheY     & 3chy  & 128   & 7 & 30 & $9.8\times 10^{-2}$ &0.093 & $2.6\times 10^{-3}$ & $8.2\times 10^{-2}$  & 3.60& 3.25 \\
ADA2h    & 1aye  & 417   & 6 & 19 & $5.9\times 10^{-6}$ &0.99 & $4.3\times 10^{-1}$ & $9.9\times 10^{-1}$  & 3.43& 2.78 \\
U1A     &  1urn  & 282 & 12 & 10 & $2.2\times 10^{-1}$ &0.17 & $1.6\times 10^{-1}$ & $9.3\times 10^{-1}$  & 3.35& 3.48 \\
ACBP    &  1aca  & 86  & 16 & 22 & $5.2\times 10^{-3}$ &0.75 & $6.7\times 10^{-2}$ & $6.3\times 10^{-1}$  & NMR &NMR \\ FKBP12  & 1fkj   & 107  & 27 & 22 & $6.7\times 10^{-4}$ &0.91 & $4.0\times 10^{-2}$ & $3.6\times 10^{-1}$  &3.11 & 2.99 \\
\hline
\end{tabular}
\vspace*{.1cm}
\caption{\small \sf
The conservation and
packing of folding nucleus residues. $N_{\mbox{prot}}$: number of
residues in the protein sequence; $N_{\mbox{seq}}$: number of
sequences; $N_\phi$: number of residues with $\phi$-value
measured. 
(a)
Correlation of participation in folding nucleus as measured by
$\phi$-value and selection pressure as measured by $\omega$. $R^2$:
the fraction of variance in the data that can be explained by the
linear regression model; $p$: the two-sided $p$-value of $t$-test for
the null hypothesis that the slope of the linear regression models is
0. 
(b) Randomization tests for assessing statistical
significance of conservation of folding nucleus residues.  The median
$\omega$ value of the folding nucleus is tested against the
distribution of the median $\omega$ value from $10^5$ random samples
containing the same number of amino acid residues as that of the
folding nucleus drawn from the same protein.  $p_{\mbox{all}}$: the
$p$-value that the folding nucleus residues are more conserved than
all other residues in the protein; $p_{\mbox{50\%}}$: the $p$-value
that folding nucleus residues are more conserved than the residue at
50\% quantile of all residues ranked by $\omega$-value.  
(c) Packing analysis of the folding nucleus and of the whole protein.
The average alpha coordination number $Z_\alpha$ for all residues in
the protein ($Z_{\alpha, \mbox{all}}$) and for residues in the folding
nucleus residues ($Z_{\alpha, \mbox{fn}}$) are listed, except for structures
determined by NMR techniques.  Protein CheY has the
highest $Z_{\alpha, \mbox{fn}}$.
}
\label{Tab:random-test}
\end{center}
\end{table*}

We use a randomization test following the approach first developed in
\cite{MirnyShakh01_JMB} to address this question. The null hypothesis
$H_0$ is that nucleus residues have equal or greater
median $\omega$ values than that of the whole protein.  That is,
folding nucleus residues are no more conserved than the whole protein
sequence.  The alternative hypothesis $H_a$ is that folding nucleus
residues have less median $\omega$ values than the whole protein
sequence and are evolutionarily more conserved.  We calculate the
median of $\omega$ values of the nucleus residues as defined in
\cite{MirnyShakh01_JMB}, and compare it with
the distribution of median of $\omega$ value in  random samples 
containing
 the same
number of residues drawn from the same protein.  As in
\cite{MirnyShakh01_JMB}, we use a  sample size of $10^5$.  The
fraction of the random samples with median $\omega$ value smaller than
that of the folding nucleus provides the $p$-value that the observed
median $\omega$-values of the folding nucleus is due to random chance. Similar to
\cite{MirnyShakh01_JMB}, we use the threshold of $p=2\%$
to decide whether evolutionary conservation of the folding nucleus
is statistically significant.  Table~\ref{Tab:random-test} shows that
$p$-value ranges between 0.26\% (CheY) and 43\% (ADA2h), but the
majority are between 4.0\% (FKBP12) and 43\% (ADA2h).  With the
exception of CheY, the null hypothesis cannot be rejected with
statistical significance at the confidence level of $p<2\%$.  That is,
except CheY, folding nuclei as defined in \cite{MirnyShakh01_JMB} in these proteins are not significantly
more conserved than the rest of the protein.

To further assess selection pressure on folding nucleus residues, we
evaluate a different null hypothesis, again using randomization test.
The null hypothesis $H_0$ now is that the folding nucleus residues
have equal or greater median $\omega$-values than the residue with
median $\omega$-value of the whole protein.  That is, folding nucleus
as defined in \cite{MirnyShakh01_JMB} are no more conserved than the
residue halfway in the rank ordered list of all residues when sorted
by estimated mean $\omega$-value.  Table~\ref{Tab:random-test} shows
that the $p$-values range from 8.2\% to 99\%.  With the criterion of
$p<2\%$, the null hypothesis cannot be rejected with statistical
significance for any of the proteins.  That is, folding nucleus for
every protein studied here is not significantly more conserved than
the residue with median $\omega$-value.

{\bf Conservation of folding nucleus of CheY.}  CheY is the only
protein among those studied here that may have a well-conserved
folding nucleus based on results of the first readomization
test. Correlation study of $\phi$-value and conservation measured by
reduced entropy also suggested that CheY protein has a well-conserved
folding nucleus \cite{Larson02_JMB}.  What are the possible reasons
for the strong conservation of folding nucleus in this protein?  It
was suggested earlier that tightly packed protein interior residues
are well conserved and these are often part of the folding nucleus
residues \cite{Privalov96,PtitsynTing99_JMB,Ptitsyn98_JMB}. We use a
parameter $z_\alpha$ recently introduced in \cite{ZhangCTL03_JCP} to
characterize protein local packing.  $z_\alpha$ is defined as
$z_\alpha \equiv n_c/n$, where $n_c$ is the number of non-bonding
atomic alpha contacts between different residues, and $n$ is the total
number of atoms.  Two atoms are in alpha contact if they are separated
by a weighted Voronoi facet which intersects with the protein
\cite{ZhangCTL03_JCP}.  $z_\alpha$ characterizes protein packing more
faithfully than other parameters such as radius of gyration
\cite{ZhangCTL03_JCP}.

We calculate $z_\alpha$ for the folding nucleus as defined in
\cite{MirnyShakh01_JMB} and for the whole protein
(Table~\ref{Tab:random-test}).  We find that the folding nucleus of
CheY has the highest $z_\alpha$ value (3.60) compared to the folding
nuclei of other proteins, whereas the whole protein $z_\alpha$ value
of CheY has similar values to other proteins.  This indicates that the
folding nucleus of CheY has significantly larger $z_{\alpha}$ than the
rest of CheY protein.  The folding nucleus of CheY is packed tighter
than folding nuclei in other proteins.  This observation can
intuitively explain the significant conservation in CheY: tight
packing in this case is accompanied by little tolerance to mutation,
since the lack of packing defects such as voids reduces the
possibility for substitution of different amino acid residues.
However, this is a rather tentative hypothesis. It is possible that
very tightly packed residues are more conserved, independent of
whether they are in folding nucleus or not.  
 It is
also possible that if results of additional experimental $\phi$-value
studies become available, the definition of the folding nucleus might
change.  To fully resolve the relationship of packing, folding, and
evolutionary conservation, more detailed additional studies are
required, which is beyond the scope of this work.

{\bf Reconstructing ancestral folding nucleus.}  The approach used in
this study can also suggest further experimental exploration of
evolution history of protein folding dynamics.  With the continuous
time Markovian model, we can reconstruct likely candidate sequences of
ancestral proteins at different evolutionary times.  Specifically,
identities of amino acid residues in the folding nucleus of ancient
ancestral proteins can be postulated.

\begin{figure}[!h]
      \centerline{
\epsfig{figure=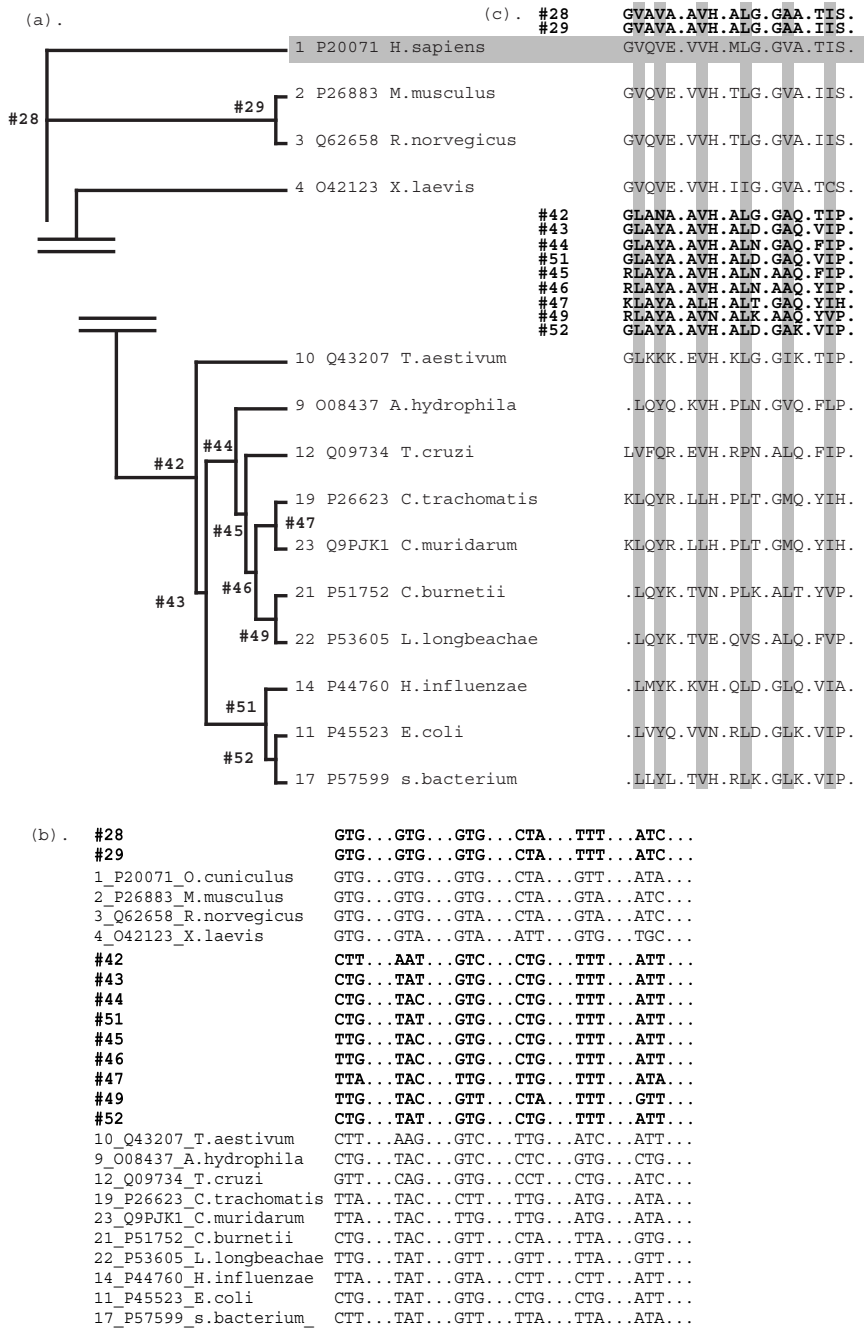,width=4.5in} } \caption{ 
\small \sf
Reconstructed ancestral
protein sequences of FKBP12 protein.  (a). The relevant part of 
the phylogenetic tree for
FKBP12 is shown.  Human FKBP12 protein from which experimental data
were obtained is shown in shadow.  (b). Multiple alignment of DNA
sequences of the folding nucleus of FKBP12 protein, including those of 
reconstructed folding nucleus of ancestral proteins. (c). Multiple alignment
of translated amino acid residue of the folding nucleus
residues identified by $\phi$-value studies (highlighted) and flanking residues.
}
\label{Fig:ancestor}
\end{figure}

As an example, we show the reconstructed residues of the folding
nuclei of FKBP12 as defined in \cite{MirnyShakh01_JMB} in
Figure~\ref{Fig:ancestor}.  The five folding nucleus residues are
VVVLVI in human FKBP12 protein.  The first residue is L in some
reconstructed ancestral genes, the second can be Y or N, the third can
be L, and the sixth can be a V instead of I.  Based on this simple
analysis, an interesting quadruplet mutagenesis study can be suggested
to experimentally test 
the folding dynamics of mutated FKBP12, where
the folding nucleus is changed. The reconstructed ancient folding
nuclei suggests a combination of residues represented by the pattern
{\bf {L\{N,Y\}L}}{\it L\/V\/}{\bf V}. Here {\bf \{N,Y\}} means either
a N or a Y residue is drawn.

The fourth residue L and fifth residue Y in all ancestral genes are
the same as that in human FKBP12, but inspection of sequences of other
extant species show that the fourth residue can be any of I, P, or V,
and the fifth can be any of I, V, L, and M.  A further interesting
experiment could be to test the folding behavior of 6-tuple mutants
with folding nucleus formed by any combination of residues represented
by the pattern {\bf L\{N,Y\}\{I,P,V\}\{I,V,L,M\}V}.  The recreated
proteins then can be assayed for folding behavior, which can be
compared with that of proteins present in extant organisms.  Such
experimental palaeobiochemistry was already envisioned by Pauling and
Zuckerkandl many years ago \cite{PaulingZuckerkandl63}, and the number
of such studies is rapidly growing
\cite{GoldingDean98_MBE,ChangDonoghue00_TEE,Adey94_PNAS,Jermann95_Nature,Chandrasekharan96_Science,DeanGolding97_PNAS}.
An in-depth study on recreating the full sequence of ancestral
proteins will require additional detailed analysis, including choosing
the most appropriate detailed evolutionary model
\cite{Schluter97_Evolution,Cunningham98_Tree,ZhangNei97_JME}.

{\bf Discussion.}  Although folding nucleus is under purifying
pressure, we fail to observe significant conservation for protein
folding nucleus residues.  Despite concerns raised in
\cite{Larson02_JMB} about the specific choices of the data in
\cite{MirnyShakh01_JMB}, we use exactly the same set of proteins, the
same definition of nuclei residues, and follow the same radomizatin
test as that of \cite{MirnyShakh01_JMB}.  It is possible that this
would bias our study towards reproducing the results of
\cite{MirnyShakh01_JMB}.  Nevertheless, our results are similar to
that of Plaxco {\it et al\/} and Larson {\it et al\/}
\cite{Plaxco00_JMB,Larson02_JMB}, and are different from that of Mirny
and Shakhnovich \cite{MirnyShakh01_JMB}.  The different conclusion of
this study and that of \cite{MirnyShakh01_JMB} is likely due to the
different evolutionary models employed, namely, the difference between
a DNA-codon based continuous-time Markov model {\it vs}.\ an implicit
evolution model implied by entropy calculation.  The conclusion that
folding nuclei residues are not conserved will likely to remain if we
were to use the data set and the definitions of folding nuclei from
reference \cite{Larson02_JMB}.  Experimental studies in barnase, SH3
domain, chymotrypsin inhibitor 2 suggest that the folding nucleus
observed in wild type protein may not be indispensable, and
alternative folding nucleus may arise if residues are mutated
\cite{MathewsFersht95_Biochem,Viguera96_NSB,Viguera02_PNAS,Itzaki95_JMB,Neira96_FD}.
Another experimental example is Im9 and Im7 proteins.  They are E
colicin-binding immunity proteins that are of the same fold with about
60\% sequence identity.  The folding of Im9 and Im7 are two-state and
three-state process, respectively.  Although these two proteins have
similar folding mechanism, $\phi$-value studies reveal that the
kinetically important residues are different
\cite{Capaldi02_NSB,Friel03_JMB}.  This is consistent with recent
simulation studies which suggest that evolution selection is more
robust for residues important for stability than for kinetic
accessibility \cite{Dokholyan02_JMB,Dokholyan02_PNAS}.  In addition,
the definition of a folding nucleus is arbitrary, because it is based
on a threshold of $\phi$ value ({\it e.g.}, $\phi \ge 0.5$)
\cite{Fersht97_COSB}.  An earlier study suggested that the critical
nucleus may be as large as $10^2$ residues, the size of a whole
protein domain \cite{BryngelsonWolynes90_Biopolymers}.  The
non-uniqueness of folding nucleus was pointed out in a study using
off-lattice model system \cite{GuoThirumalai96_FD}.  The role of
protein structure in folding is discussed from the viewpoint of
small-world connections in \cite{Vendruscolo02_PRE}.  Recent
computational studies based on exact enumerable lattice models using
master equation showed that there are remarkable heterogeneity in
structural contacts underlying macroscopic two-state folding kinetics
of model G\={o} protein \cite{Ozkan1_NSB,Ozkan02_PS}.  The kinetic
barrier was shown to result from a reduced number of microroutes near
the bottom of the folding funnel \cite{Ozkan1_NSB,Ozkan02_PS}.  If
these studies portray accurately the microscopic picture of the
folding process, there are likely to be many different native contacts
that form folding nuclei for different folding pathways in the free
energy landscape. It is reasonable to expect that a large subset of
residues are capable of providing critical native contacts, and these
contacts vary for different microscopic folding pathways.  The roles
of these residues in folding are largely interchangeable, and this may
be reflected in the lack of extraordinarily strong purifying selection
pressure in the current set of folding nucleus residues characterized
by $\phi$-value studies.

In summary, we use a continuous time Markovian model
\cite{GoldmanYang94_MBE} and apply a maximum likelihood estimator
developed in \cite{PAML} to study the evolution of protein folding
dynamics.  We examine the coding DNA sequences rather than amino acid
residue sequences, and assess selection pressure by estimating the
ratio $\omega$ of nonsynonymous {\it vs.}\ synonymous codon
substitution rate. The position specific rate ratio is used to
distinguish substitutions fixed by evolution and by chance.  We found
that folding nucleus residues experience purifying selection pressure,
but they are not significantly more conserved than the rest of the
residues of the whole protein.  The only exception is CheY protein,
where the folding nucleus is significantly more conserved.  This may
be due to extraordinarily tight packing, which is reflected by the
high alpha coordination number $Z_{\alpha}$.  
Results described here provides another
confirmation that evolution does not preserve kinetically important
residues, which has been a subject of debate in literature
\cite{Plaxco00_JMB,MirnyShakh01_JMB,Larson02_JMB}.
We further suggest
exploratory palaeobiochemical studies testing the evolution of protein
folding dynamics.

{\bf Acknowledgment.}
We thank Drs.\ Xun Gu, Robie Mason-Gamer, and Clare Woodward for
helpful discussions, Dr.\ Ziheng Yang for generous help in using {\sc
Paml}.  This work is supported by grants from National Science
Foundation (CAREER DBI0133856 and DBI0078270) and National Institute
of Health (GM68958).

\bibliography{evo,nigms,svm,array,prop,pack,prf,lattice,bioshape,liang,pair,potential,career,design,pd02,evo}
\bibliographystyle{unsrt}
\end{document}